\documentclass[fleqn,usenatbib]{mnras}

\usepackage{newtxtext,newtxmath}

\usepackage[T1]{fontenc}
\usepackage{ae,aecompl}

\usepackage{graphicx}	
\usepackage{amsmath}	
\usepackage{amssymb}	

\title[Prediction of Emission Line Galaxy Surveys] {Prediction of H$\alpha$ and [OIII] Emission Line Galaxy Number Counts for Future Galaxy Redshift Surveys}

\author[Z. Zhai et al.]{
Zhongxu Zhai,$^{1}$\thanks{E-mail: zhai@ipac.caltech.edu}
Andrew Benson,$^{2}$
Yun Wang,$^{1}$
Gustavo Yepes,$^{3,4}$
Chia-Hsun Chuang$^{5}$
\\
$^{1}$IPAC, California Institute of Technology, Mail Code 314-6, 1200 E. California Blvd., Pasadena, CA 91125 \\
$^{2}$Carnegie Observatories, 813 Santa Barbara Street, Pasadena, CA 91101 \\
$^{3}$Departamento de F\'isica Te\'{o}rica, M\'{o}dulo 8, Facultad de Ciencias, Universidad Aut\'{o}noma de Madrid, 28049 Madrid, Spain \\
$^{4}$CIAFF, Facultad de Ciencias, Universidad Aut\'{o}noma de Madrid, 28049 Madrid, Spain \\
$^{5}$Kavli Institute for Particle Astrophysics and Cosmology, Stanford University, 452 Lomita Mall, Stanford, CA 94305
}
\date{Accepted XXX. Received YYY; in original form ZZZ}

\pubyear{2015}

\hypersetup{draft}
\begin{document}
\label{firstpage}
\pagerange{\pageref{firstpage}--\pageref{lastpage}}
\maketitle

\begin{abstract}

We perform a simulation with Galacticus, a semi-analytical galaxy formation model, to predict the number counts of H$\alpha$ and [OIII] emitting galaxies. With a state-of-the-art N-body simulation, UNIT, we first calibrate Galacticus with the current observation of H$\alpha$ luminosity function. The resulting model coupled with a dust attenuation model, can reproduce the current observations, including the H$\alpha$ luminosity function from HiZELS and number density from WISP. We extrapolate the model prediction to higher redshift and the result is found to be consistent with previous investigations. We then use the same galaxy formation model to predict the number counts for [OIII] emitting galaxies. The result provides further validation of our galaxy formation model and dust model. We present number counts of H$\alpha$ and [OIII] emission line galaxies for three different line flux limits: $5\times10^{-17}$erg/s/cm$^{2}$, $1\times10^{-16}$erg/s/cm$^{2}$ (6.5$\sigma$ nominal depth for WFIRST GRS), and $2\times10^{-16}$erg/s/cm$^{2}$ (3.5$\sigma$ depth of Euclid GRS). 
At redshift $2<z<3$, our model predicts that WFIRST can observe hundreds of [OIII] emission line galaxies per square degree with a line flux limit of $1\times10^{-16}$erg/s/cm$^{2}$. This will provide accurate measurement of large scale structure to probe dark energy over a huge cosmic volume to an unprecedented high redshift. Finally, we compare the flux ratio of H$\alpha$/[OIII] within the redshift range of $0<z<3$. Our results show the known trend of increasing H$\alpha$/[O III] flux ratio with H$\alpha$ flux at low redshift, which becomes a weaker trend at higher redshifts. 

\end{abstract}


\section{Introduction}

Understanding the nature of the cosmic acceleration is one of the prominent questions in modern cosmology. This phenomenon was first discovered through the observation of type Ia supernovae (SNe Ia, \citealt{Riess_1998, Perlmutter_1999}). Its unknown cause is known as dark energy, which could be a negative-pressure component in the universe, or the modification of general relativity. The simplest model of dark energy is the cosmological constant. However, its value inferred from cosmological observations differs from the prediction of quantum field theory by $\sim$ 120 orders of magnitude (\citealt{Weinberg_1989}). This discrepancy has motivated theorists to develop many different models of dark energy, as well as alternative theories of gravity. However, current cosmological measurements are insufficiently accurate to rule out many of these models. For instance, galaxy redshift surveys like BOSS (\citealt{Dawson_BOSS}), eBOSS (\citealt{eBOSS_Dawson}), and WiggleZ (\citealt{Drinkwater_2010}), are designed to measure the expansion history of the universe and the growth rate of the large scale structure through the baryon acoustic oscillation (BAO) and redshift space distortions (RSD). These measurements can provide important information to constrain dark energy and modified gravity models (\citealt{Wang_2008a, Wang_2008b, Wang_2008c, Wang_2010}). However, the possible parameter space is still large and therefore more powerful surveys with larger area and deeper fields are needed.

Examples of such future surveys include LSST (\citealt{LSST-sciece-book}), DESI (\citealt{DESI_2016}), 4MOST (\citealt{Jong_2016}), and PFS (\citealt{Takada_2014}) from the ground, and two space missions: ESA's Euclid (\citealt{Laureijs_2011, Laureijs_2012}) and NASA's WFIRST (\citealt{Green_2012, Dressler_2012, Spergel_2015}). Euclid and WFIRST will use near-IR grism spectroscopy to measure tens of millions of emission line galaxies (ELGs) at intermediate redshifts. The resulting data can provide measurement of the BAO signal and RSD effect (\citealt{Glazebrook_2005}), as well as galaxy evolution and star formation histories. The main target of these surveys will be H$\alpha$ and [OIII] $\mathrm{\lambda}5007$ emission line galaxies (ELGs). The Euclid survey will observe the H$\alpha$ ELGs in the redshift range $0.9<z<1.8$, while WFIRST has a slightly different redshift coverage $1.0\la z \la 2.0$. The [OIII] ELGs for both surveys will be targeted at $z>2.0$. The Euclid and WFIRST surveys are designed to have similar but complementary strategies in the sense that Euclid has a much wider survey area and relatively shallow H$\alpha$ flux limit, while WFIRST has smaller area and greater depth. This will enable a cross check of the systematic effects and lead to more robust constraints on dark energy.

One of the important tasks for these future surveys is to optimize the survey designs and evaluate their performance in constraining dark energy. Among the quantities that can shape the capability of the future surveys, the number density of the target galaxies (as a function of redshift) is of critical importance. Therefore, realistic predictions are necessary to prepare for future surveys like Euclid and WFIRST. This can be done through numerical simulations to produce highly realistic synthetic galaxy mock catalogs. In order to do so, it is necessary to populate high-resolution N-body simulations with galaxies of particular types. 

This can be achieved in several ways, including using a halo occupation distribution (HOD, \citealt{HOD_Weinberg, Peacock_2000, Seljak_2000, Benson_2000, Martin_2001, Cooray_2002}) approach, direct hydro-dynamical simulation (\citealt{Springel_2001, Pearce_2001, Somerville_2015}) with modeling of nebular emission lines (\citealt{Hirschmann_2017, Hirschmann_2019}), or by using semi-analytical galaxy formation models (SAM, \citealt{White_1991, Kauffmann_1993, Somerville_1999, Cole_2000, Bower_2006, Orsi_2010, Benson_2012, Stothert_2018, Izuierdo-Villalba_2019}). In this paper, we present our work using the SAM approach, which uses parameterized prescriptions to describe the various astrophysical processes governing the formation of galaxies within the cosmic large-scale structure. Our choice of using a SAM for this work is driven by the fact that the empirical HOD approach depends on observations which are not available for the WFIRST mission, and the hydro-dynamical simulation is far too slow to generate sufficient numbers of galaxies to populate such a large volume with the required resolution. 

However, we note that there are several investigations attempting to estimate the number densities of target galaxies based on current observations, which can provide useful input for Euclid and WFIRST. For instance, \citet{Colbert_2013} and \citet{Mehta_2015} measure the number density of H$\alpha$ emitters based on the data collected from Wide Field Camera 3 (WFC3) Infrared Spectroscopic Parallels survey (WISP; \citealt{Atek_2010, Atek_2011}).  Although the survey area of WISP is small and the redshift coverage does not match that of Euclid or WFIRST completely, the similarity in the observational mode (slitless grism) and the larger overlap in wavelength coverage make the WISP results highly relevant for such future surveys. \citet{Pozzetti_2016} uses empirically motivated models to describe the H$\alpha$ luminosity function based on the current observational data. In addition, \citet{Colbert_2013} and \citet{Valentino_2017} estimate the number densities of [OIII] ELGs. The redshift ranges of these estimates are lower than the redshift ranges spanned by Euclid and WFIRST, but they can nevertheless provide a low-redshift anchor for the predictions presented in this work. The analysis in this paper is similar in approach to that of \citet{Merson_2018} which provides a detailed prediction for the H$\alpha$ ELGs, and the redshift dependence at different flux limits for Euclid and WFIRST. We will focus on the [OIII] ELGs in this paper, presenting the first model predictions for their number counts with redshift ranges and flux limits relevant to Euclid and WFIRST. Since we base our calculation on a state-of-the-art N-body simulation, which provides more accurate and better statistics for the relevant observables, we present updated number counts of H$\alpha$ ELGs as well.

Our paper is organized as follows: in Section 2, we introduce the galaxy formation model and its calibration with current observations. Section 3 presents our calculation and prediction for the H$\alpha$ emission line galaxies. The results for [OIII] are shown in Section 4. Finally we discuss and conclude in Section 5.

\section{Galaxy formation model and calibration}

In this work, we apply Galacticus (\citealt{Benson_2012}) --- a semi-analytical galaxy formation model to perform the prediction and analysis \footnote{We use the version 0.9.6 of Galacticus, which is publicly available at \url{https://github.com/galacticusorg/galacticus}. The hash ID for this version is 95b99550c9cc85ceb0ea8d0f63a2c8a9c7a32fe8.}. This section describes the details of the model, including the input merger tree catalog and the model calibration.

\subsection{Galaxy formation model: Galacticus}

Galacticus forms and evolves model galaxies using an approach similar to other SAMs. The input to the model is a set of hierarchical merger trees of dark matter halos, which can be constructed by the Press-Schechter formalism or through cosmological N-body simulations. The baryonic processes related to galaxy formation and evolution within these dark matter halos are parameterized by a set of coupled ordinary differential equations (ODEs). These include the rate of gas cooling, the star formation rate, the chemical enrichment of the stellar and gaseous component, feedback processes from supernovae and active galactic nuclei, the evolution of supermassive black holes and so on. The output from Galacticus is a set of properties of the galaxies, including the redshift, stellar mass, size, metallicity, morphology, star formation history and so on. In addition, Galacticus can also calculate the spectral energy distribution (SED) for each galaxy, given models for the stellar initial mass function (IMF) and a set of simple stellar population spectra. An analysis of the SEDs of these galaxies, as well as their evolution with redshift will be presented in a future paper.

The emission line luminosity of the galaxies from Galacticus is computed using the CLOUDY photoionisation code (\citealt{Ferland_2013}). The details of the method are fully described in \citet{Merson_2018}. The key step is to generate and interpolate tabulated libraries of emission line luminosities using CLOUDY as a function of the number of ionizing photons for various species (H I, He I and O II), the metallicity of the interstellar medium (ISM), the hydrogen gas density and the volume filling factor of H II regions, which can be computed for the galaxies from Galacticus. 

Galacticus is designed to be highly modular and flexible such that new physical ingredients can be easily added in the model. In our analysis, we adopt the stable branch v0.9.6. The parameters of the models are described in more detail in Section \ref{sec:calibration}.

Galacticus, like other SAMs, can be calibrated to reproduce various statistics of a galaxy population. The comparison with observational facts has enabled such galaxy formation models to provide useful evidence for the underlying physics. This includes the examination of galaxy stellar mass function, star formation rate density, galactic conformity, gas-phase metallicity and so on, see e.g. \citealt{Power_2010,  Lu_2011, Kauffmann_2013, Fu_2013, Somerville_2015b, Knebe_2015, Croton_2016} and references therein. In this analysis, we focus on the reproduction of luminosity function covered in a wide redshift range and the implication for future galaxy surveys.

\subsection{Merger tree catalogs}

The dark matter halo merger trees used as input for Galacitus are extracted from one of the latest N-body simulations, UNIT\footnote{http://www.unitsims.org} (\citealt{Chuang_2019}), which is designed to focus on characterizing statistics relevant to emission line and luminous red galaxies in large galaxy surveys. 
UNIT adopts suppressed variance methods and consists of a suite of fully N-body simulations ({\sc Gadget}, \citealt{2005MNRAS.364.1105S}) and particle mesh simulations (FastPM, \citealt{Feng:2016yqz}). In particular, we use the full N-body calculation (i.e. {\sc Gadget}) in our analysis. The simulation is started from redshift $z=99$ and is run to $z=0$. 

It assumes a spatially flat $\Lambda$CDM model with the parameters:  $\Omega_{m} = 0.3089$, $h=H_{0}/(\text{km s}^{-1} \text{Mpc}^{-3})/100 = 0.6774$, $n_{s} = 0.9667$, and $\sigma_{8} = 0.8147$, consistent with the Planck 2016 measurement (\citealt{Planck_2016}). The simulation contains $4096^3$ particles with a box-size of $1h^{-1}\text{Gpc}$. The resulting particle mass is $\sim10^{9}h^{-1}M_{\sun}$. For more details of the simulation, we refer the reader to \citet{Chuang_2019}.

Dark matter halos are identified using the publicly available ROCKSTAR halo finder (\citealt{Behroozi_2013a}), and merger trees are constructed using the Consistent Trees software (\citealt{Behroozi_2013b}). Due to the high resolution of this simulation, the total number of the merger trees is approximately 160 million.

In order to forecast the number counts of H$\alpha$ and [OIII] emitting galaxies, we must first build a light cone catalog. The Galacticus model implements light cone construction with the method from \citet{Kitzbichler_2007}. The resulting catalog in our analysis has a survey area of 4 deg$^{2}$ in the redshift range $0<z<3$. This redshift range is chosen to match the observations of the WFIRST and Euclid missions, as well as follow up investigations for the distribution and properties of the SEDs of the galaxies.

\subsection{Calibration with H$\alpha$ luminosity function} \label{sec:calibration}

\begin{figure*}
\begin{center}
\includegraphics[width=18cm]{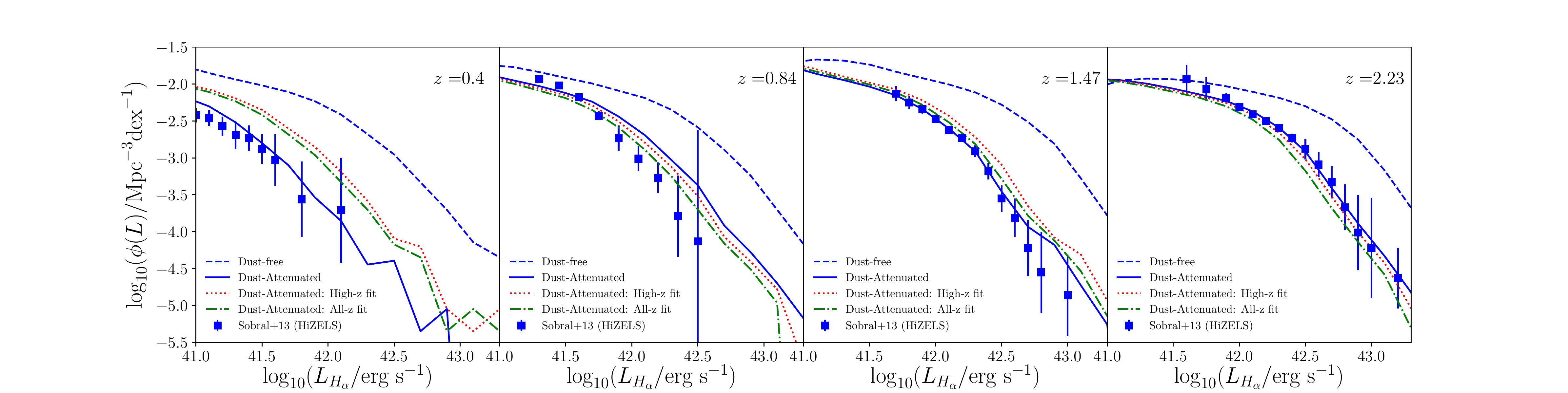}
\caption{Calibration of the galaxy formation model: comparison of luminosity functions from HiZELS (\citealt{Sobral_2013}) and our calibrated model. The redshifts at which the luminosity function are calculated are chosen to match the WFIRST and Euclid observation strategy, as shown in the upper-right corner of each panel. The dashed line indicates the dust-free result from the Galacticus model, while the solid line denotes the dust-attenuated result which is obtained with the best-fit optical depth at the particular redshift. The green dot-dashed line is dust-attenuated by assuming a constant optical depth fitted across all the redshifts of HiZELS observation, while the red dotted line corresponds to the optical depth fitted with 3 high redshift measurements.}
\label{fig:Hizels}
\end{center}
\end{figure*}

\begin{figure}
\begin{center}
\includegraphics[width=8.5cm]{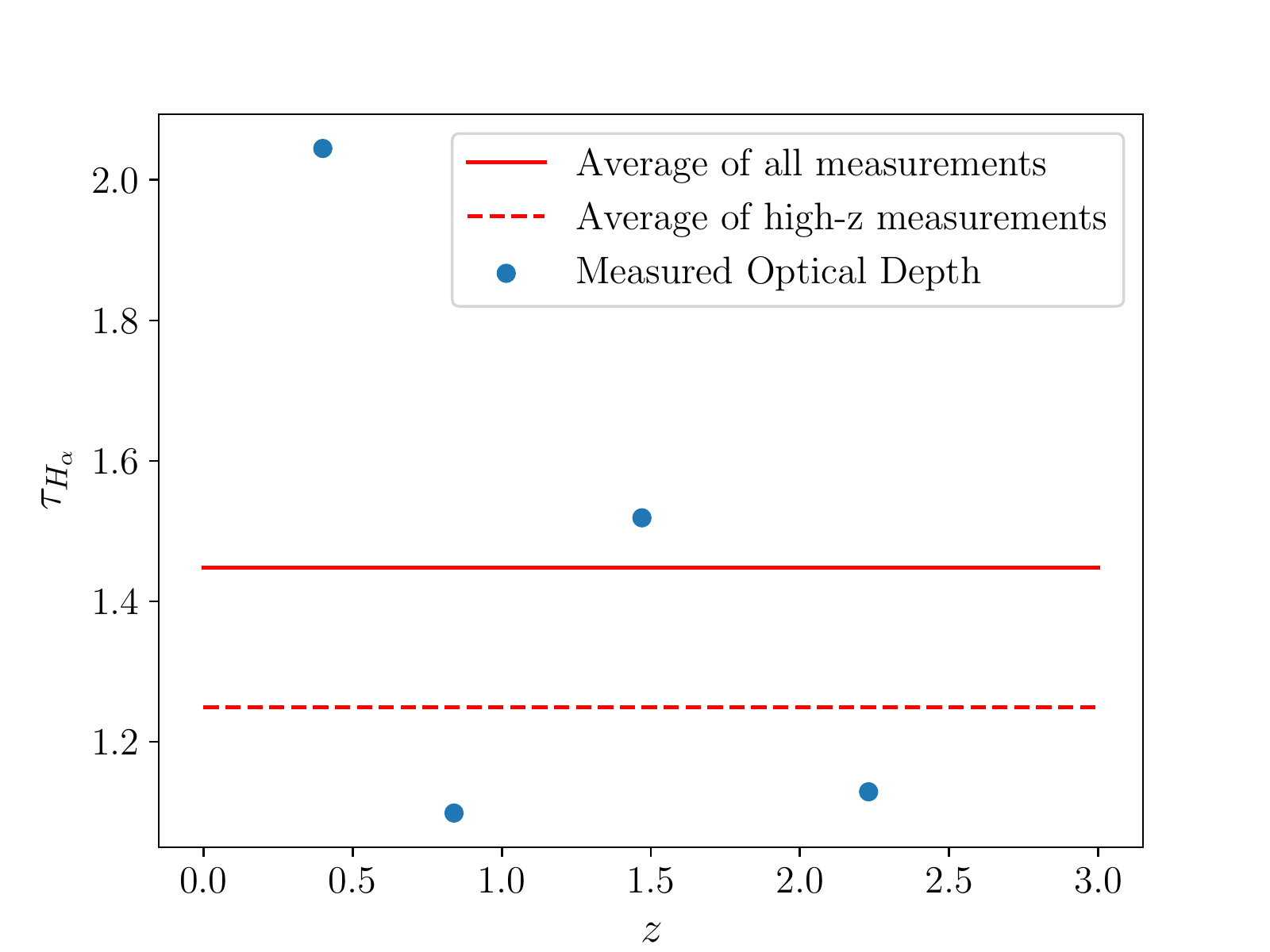}
\caption{Calibration of the dust model: optical depth of H$\alpha$ emission line luminosity measured to match the observed luminosity function. The horizontal solid and dashed lines are average of all 4 measurements and 3 high-z measurements respectively.}
\label{fig:tau}
\end{center}
\end{figure}

As with all semi-analytical models, Galacticus must be calibrated to reproduce some statistics of the galaxy population in the local universe, or at higher redshifts where relevant observational data is available. Since the UNIT simulation has a new set of cosmological parameters and mass resolution, the parameters of Galacticus in the previous studies (such as \citealt{Merson_2018}) are not appropriate. Therefore we must find a new calibration of model parameters for the UNIT simulation and verify that Galacticus can then reproduce the statistics of interest. 

In this analysis, we do not limit ourselves to the local universe to calibrate the model. Instead, we focus on observational data in the relevant redshift range for Euclid and WFIRST. In particular, we choose the high redshift H$\alpha$ luminosity function (LF) measurements by the ground-based narrow-band High-z Emission Line Survey (HiZELS, \citealt{Geach_2008, Sobral_2009, Sobral_2013}), from Table 4 of \cite{Sobral_2013}, as shown in Figure \ref{fig:Hizels}. Note that in their Table 4, \cite{Sobral_2013} corrected the H$\alpha$ luminosity by 1 mag to account for dust extinction; thus we need to undo that by subtracting 0.4 from log$_{10}L$, in order to obtain the observed LF without dust extinction correction, to compare with predictions by Galacticus assuming a dust model. Dust extinction cannot be determined from the observations, thus it is most useful to compare dust obscured LF, as our methodology is to vary the dust model in Galacticus to match the observed LF.

The luminosity function, as well as many other properties of galaxies, can depend on the many astrophysical processes and parameters modeled in Galacticus. In our calibration, we allow the tunning of these processes: cooling of gas, star formation process, SNe feedback, black hole feedback, galaxy merging and morphology, metal yield and dust attenuation. The parameter input file of this work is made available online as a supplementary dataset.

Given the high-dimensional parameter space and the size of the simulation, it is not practical to explore the parameter space with a Monte Carlo Markov Chain (MCMC)-like approach. A possible solution is the emulator technique as presented in \citet{Bower_2010} to quickly evaluate the model prediction at any point in the parameter space. Due to large uncertainties in the measurement of the luminosity function, and our goal of searching for a reasonable parameter set instead of accurate statistical inference, we adopt a simpler method in the calibration. We first choose a subsample of the merger trees that can represent the whole simulation, and then use a Latin hypercube method to define hundreds of models that uniformly sample the parameter space. We note that this method has been widely used in the design of cosmological simulations (\citealt{Heitmann_2009, Garrison_2018, DeRose_2019}) and building emulators for statistics of dark matter (halos) and galaxies (\citealt{Heitmann_2010, Lawrence_2010, McClintock_2019, Zhai_2019}). We then use Galacticus to generate galaxy populations for each such parameter set. From the resulting galaxy output we calculate the H$\alpha$ luminosity functions of the galaxies at the same redshifts as HiZELS observations and evaluate a $\chi^{2}$ measure of the goodness of fit of the Galacticus model to the data:
\begin{equation} \label{eq:chi2}
\chi^{2} = (\phi_{\mathrm{obs}, i, z}-\phi_{\mathrm{pre}, i, z}) C^{-1} (\phi_{\mathrm{obs}, i, z}-\phi_{\mathrm{pre}, i, z}), 
\end{equation}
where $\phi$ is the luminosity function, the subscript ``pre'' and ``obs'' refer to model prediction and HiZELS observation respectively, ``$i$'' and ``$z$'' denote the luminosity and redshift bin respectively, and $C$ is the covariance matrix of the measurement. We then search for the model that has the minimum $\chi^{2}$ and choose its parameter set as our calibrated model. 

An important component in the calibration process is the dust attenuation model. \citet{Merson_2018} applied three different dust models in their analysis: \citet{Ferrara_1999}, \citet{Charlot_2000}, and \citet{Calzetti_2000}. The prediction of the H$\alpha$ number counts based on these three dust models are roughly consistent, with the \citet{Calzetti_2000} model giving predictions that are in best agreement with observations. Thus we adopt the \citet{Calzetti_2000} dust model in our analysis, since it is also the most economic computationally.

The \citet{Calzetti_2000} model is an empirical approach and has been widely used in observational analyses. We express the attenuation of luminosity by dust in this model as
\begin{equation}
L_{\mathrm{att}} = L_{0}10^{-0.4C(\lambda)A_{v}},
\end{equation}
where $L_{\mathrm{att}}$ is the dust attenuated luminosity and $L_{0}$ is the dust-free, intrinsic luminosity, $C(\lambda)$ is related to the extinction curve and depends only on wavelength, $A_{v}$ is a free parameter to be determined by data. Once the value of $A_{v}$ is determined, this dust model is complete and applicable to all emission lines. We note that the luminosity function measurements from HiZELS cover a wide redshift range, so it is possible that a single value of $A_{v}$ is only a rough approximation to represent the dust attenuation. In particular, Galacticus has implemented an analysis module to compare the output galaxies with the given observational constraints which in this paper, is the dust-attenuated luminosity function from HiZELS. Therefore the parameter in the dust model is also a parameter in our Latin hypercube design and can depend on redshift. So we choose the physical parameters at redshift $z=1.47$ that can reproduce the observed HiZELS luminosity function measurement at the same redshift with a minimum of $\chi^{2}$, as our optimized model to generate the mock galaxies. We also note that this redshift is the most relevant for the WFIRST and Euclid observing strategies. In order to calibrate the dust model within the whole HiZELS redshift range, we model the effect of dust on the luminosity function through the optical depth at H$\alpha$ wavelength as defined by \citet{Merson_2018}

\begin{equation}
\tau_{H{\alpha}} = -\ln\left({\frac{L_{H\alpha}^\mathrm{att}}{L_{\mathrm{H}\alpha}^{0}}}\right),
\end{equation}
where the $L_{H\alpha}$ is the H$\alpha$ luminosity, and the superscript ``att'' and ``0'' denote the dust-attenuation and dust-free respectively. With the Galacticus galaxies, we apply this model to find the dust-attenuated H$\alpha$ luminosity by searching for a value of $\tau_{H\alpha}$ that results in a  prediction for the H$\alpha$ luminosity function closest to the HiZELS measurement at each redshift. This can be done by a simple $\chi^{2}$ computation as in Eq. (\ref{eq:chi2}) and the result is shown in Figure \ref{fig:tau}. The dust-attenuated luminosity function from our calibrated Galacticus model is shown in Figure \ref{fig:Hizels} and is consistent with the HiZELS measurements.

Figure \ref{fig:tau} shows that the optical depth varies significantly with redshift. In order to assess the effect of the dust model on the resulting prediction of the galaxy distribution, we assume a constant optical depth fit to measurements at all 4 redshifts and only at the 3 highest redshifts, and compare the prediction of the galaxy number counts. With these constrained optical depths, we show the dust attenuated luminosity function in Figure \ref{fig:Hizels}. Since the average optical depth of all 4 measurements is higher than the average value of 3 high-z data points, the resulting luminosity function is lower, but can provide a reasonable fit to the HiZELS observations in the WFIRST redshift window. On the other hand, the optical depth fitted with the three high-z measurements (dotted red line) shows some deviation compared with HiZELS, we can see that this dust model can give more consistent result with WISP measurement in the next section.
In addition, the optical depth is related to the parameter $A_{v}$ in the \citet{Calzetti_2000} model through
\begin{equation}
A_{v} = -\log_{10}(-\exp(\tau_{H\alpha}))/(0.4C(\lambda=H\alpha)).
\end{equation}
We then apply this value of $A_{v}$ in the dust model to obtain the luminosities of the other emission lines. In other words, at a particular redshift, the dust extinction still obeys the \citet{Calzetti_2000} law to model the wavelength dependence. We also apply this dust model in the following calculation for the prediction of the number counts of H$\alpha$ and [O III] emission line galaxies. 
We note that the dust correction in our calibration is only an approximate method, especially the assumption of constant optical depth across redshift. The uncertainty in the following calculation based on this implicitly absorbs uncertainties from other sources, such as any intrinsic excess of emitters in the SAM with respect to observations, or the difference in cosmologies assumed by HiZELS and the N-body simulation we use. Other statistics of the resulting galaxy catalog are presented in the appendix.

\section{Predictions for H$\alpha$ emission line galaxies}

With the above calibrated Galaticus and dust models, we first estimate the number density of the H$\alpha$ emission line galaxies and compare with previous studies. 

\begin{figure}
\begin{center}
\includegraphics[width=8.5cm]{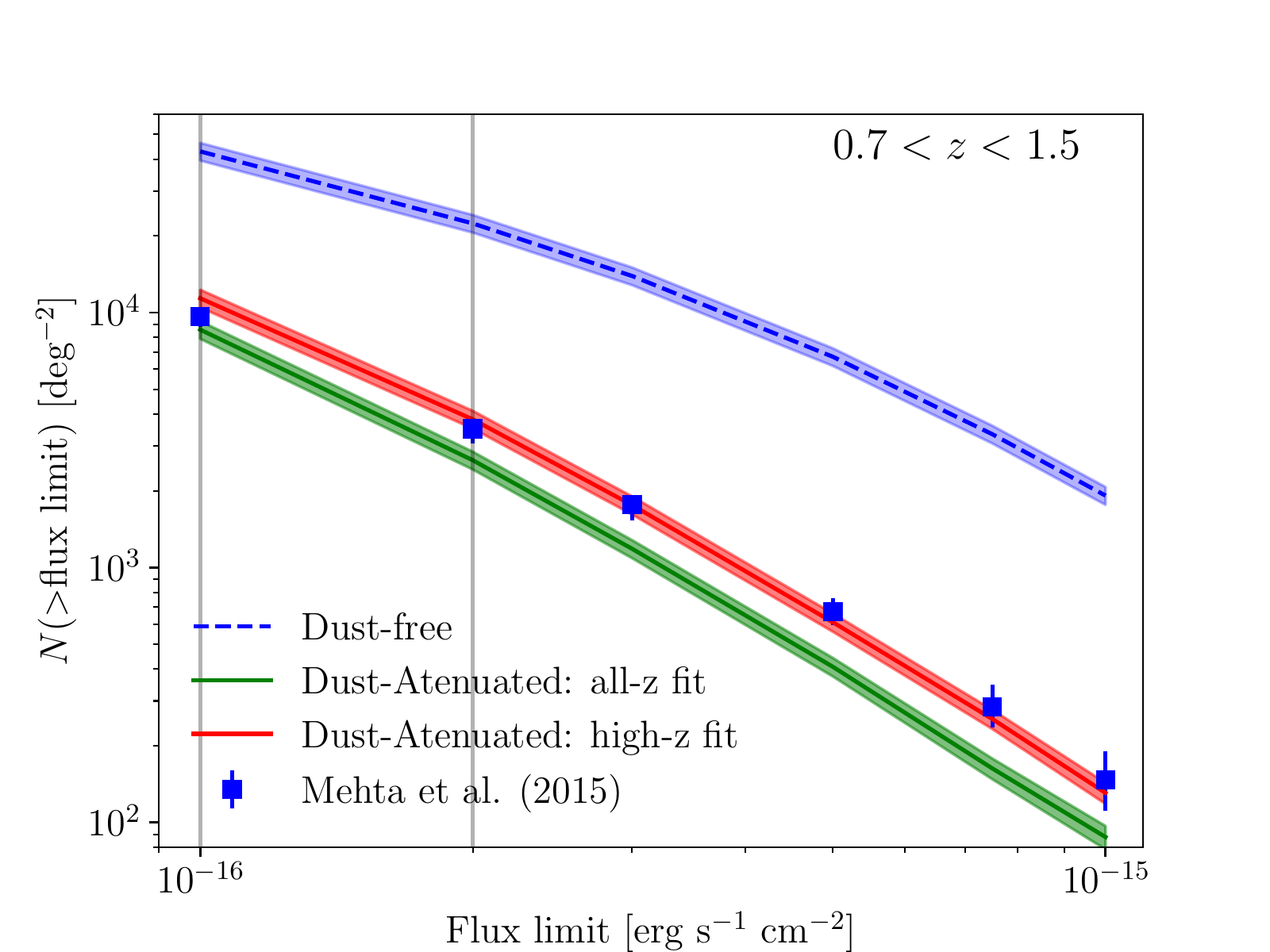}
\caption{The distribution of the cumulative H$\alpha$ flux counts as predicted by Galacticus. The dashed line is the dust-free result for comparison, while the solid lines adopt our calibrated dust models with optical depth fitted at all redshift or just high redshift respectively. The uncertainty of the measurement is obtained by the jackknife resampling method combined with the cosmic variance. Note that the cosmic variance is estimated to match the redshift range $0.7<z<1.5$, which is approximately a 7.9\% variation. The squares with errorbars are measurements from WISP (\citealt{Mehta_2015}) in the same redshift range. The vertical gray lines indicate the flux limits of 1 and 2$\times10^{-16}\text{ergs}/\text{s}/\text{cm}^2$, corresponding to $6.5\sigma$ nominal depth for WFIRST and $3.5\sigma$ depth for Euclid respectively.}
\label{fig:Hcounts}
\end{center}
\end{figure}

Figure \ref{fig:Hcounts} shows the number counts of the H$\alpha$ emitters as a function of flux limit within the redshift range $0.7<z<1.5$, for both the dust free and dust attenuated luminosity respectively. The significant difference between the dust free and dust attenuated results shows that the dust model is necessary in the modeling of galaxy distribution. The uncertainty of the measurements in our analysis are obtained by subsampling. In particular, we split the galaxies according to their right ascension and declination into 25 subregions, each has an area of 0.16 deg$^{2}$ and contains a similar number of galaxies. We then compute the number counts by excluding one of the subregions and estimate the uncertainties using the jackknife approach. Given the 4 deg$^{2}$ lightcone, we also estimate the cosmic variance using the fitting formula as presented in \citet{Driver_2010}. For a WFIRST like survey with redshift $1<z<3$, the cosmic variance is at 4.9\% percent level for the estimated number counts. We then add these two error budgets in quadrature as our total uncertainty. The final uncertainties estimated in this way are shown as the shaded area in the figure. For comparison, we also show the estimate from the WISP survey by \citet{Mehta_2015} in the same redshift range. Although the WISP-based analysis has a total area of only approximately 0.051 deg$^{2}$, it still carries important information as a reference. 

When we apply the dust model fit to the optical depth at high redshifts, the prediction from our analysis is consistent with the WISP measurements. However, when we adopt the average value of the optical depth that can fit the HiZELS observations across all redshifts, the number counts are lower than the WISP results, especially at higher flux limit. With the flux limit of a WFIRST-like survey of $1\times10^{-16}$erg s$^{-1}$cm$^{-2}$, the mean number counts that we have predicted are higher than the WISP measurements by $20-30\%$, which is a similar amount of variation given the survey area. In addition, the consistency between our prediction and the WISP measurement also implies consistency with the previous investigation in \citet{Merson_2018}, thereby validating our calibration of the galaxy formation model. 

\begin{figure}
\begin{center}
\includegraphics[width=8.5cm]{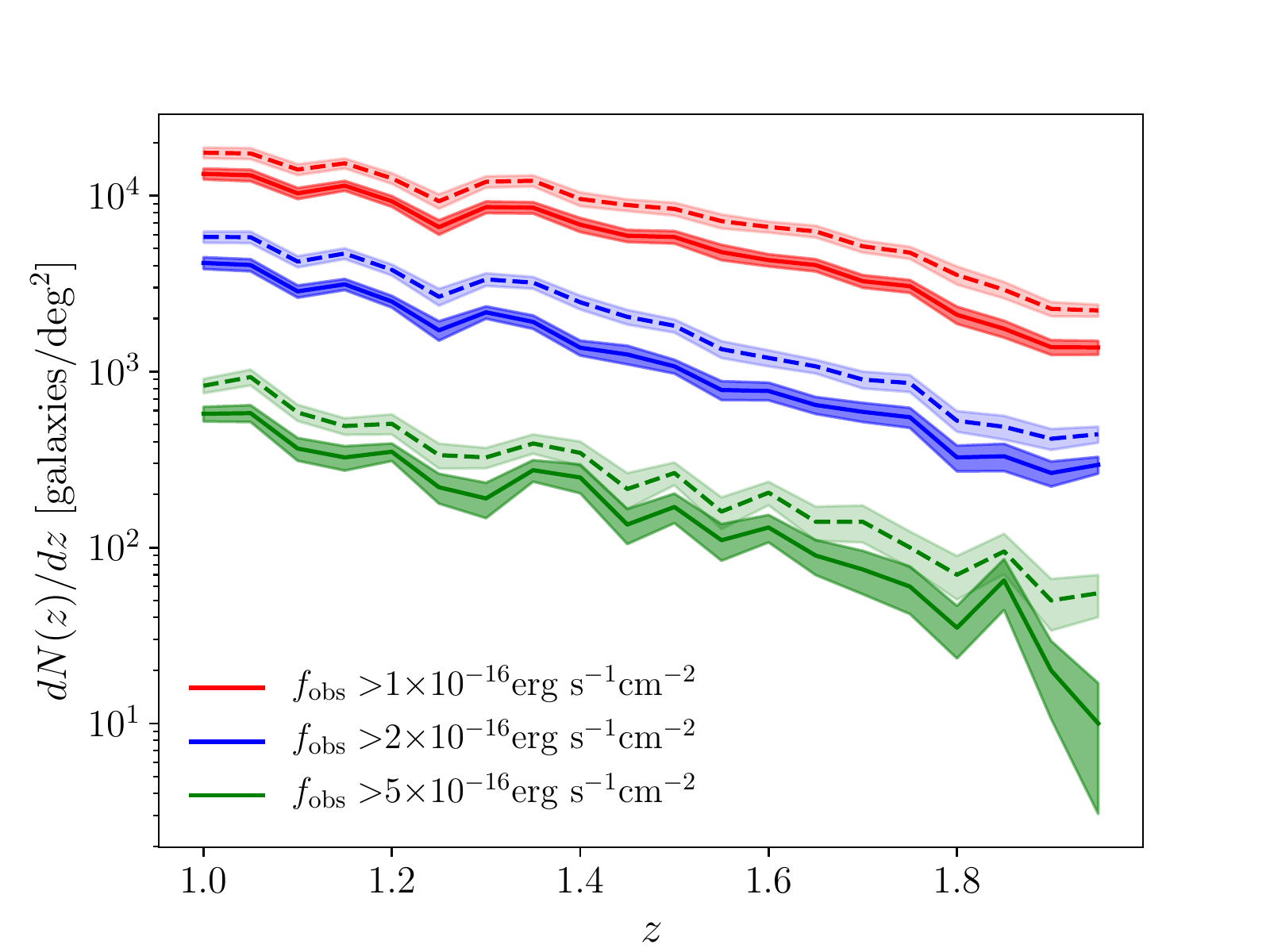}
\caption{Predictions for the galaxy redshift distribution of the H$\alpha$ emitting galaxies. The results are shown for different flux limits as indicated in the legend, which correspond to WFIRST and Euclid survey strategies. The shaded area shows $1\sigma$ uncertainty estimated by jackknife resampling and cosmic variance. The solid lines are obtained with dust model fitted with the entire HiZELS redshifts, while the dashed lines are from the high redshift fit only.}
\label{fig:Hredshift}
\end{center}
\end{figure}

In Figure \ref{fig:Hredshift} and Table \ref{tab:Halpha}, we present the prediction for the redshift distribution of H$\alpha$ number counts for different flux limits and different dust models. Note that in Figure \ref{fig:Hredshift}, we have added a brighter flux limit case for comparison with Euclid and WFIRST, to demonstrate the importance of a sufficiently faint flux limit to probe the higher redshifts. In Table \ref{tab:Halpha}, we have added a fainter flux limit case to illustrate the gain in deeper surveys. The uncertainties are obtained by the same jackknife resampling method and cosmic variance as described above. We restrict the prediction to the redshift range of $0.5<z<2.0$ since this spans the entire redshift range for H$\alpha$ ELGs relevant to both WFIRST and Euclid. The results show an overall decline of H$\alpha$ number density with redshift, regardless of the flux limit, but the rate of decline increases at brighter fluxes. This is a reflection of the LF of H$\alpha$ ELGs, which declines sharply at the bright end (see Fig.\ref{fig:Hizels}). The apparent peaks and troughs in the redshift distribution are likely caused by sample variance.

As expected based on the H$\alpha$ LF, the redshift distribution of H$\alpha$ number counts is very sensitive to the line flux limit. WFIRST has a flux limit of $1\times10^{-16}$erg s$^{-1}$cm$^{-2}$ (6.5$\sigma$), while Euclid has a flux limit of $2\times10^{-16}$erg s$^{-1}$cm$^{-2}$ (3.5$\sigma$). From Fig.\ref{fig:Hredshift}, we can see that WFIRST can observe more than twice as many galaxies than Euclid due to its fainter flux limit. 
This will have a direct consequence for data analysis. 
The shot noise in the galaxy clustering measurement is set by the inverse of $nP$ (with $nP\sim 1$ being the "rule of thumb" survey design goal), where $n$ is the space density of the target galaxy and $P$ is the amplitude of the power spectrum in the region of interest (\citealt{DESI_2016}). Therefore WFIST will have large enough number density to enable better modeling of systematic uncertainties, while Euclid will achieve high statistical precision due to its larger survey volume.

\begin{table*}
\centering
\caption{Predicted redshift distribution of the number counts of H$\alpha$-emitting galaxies, ${\rm d}N/{\rm d}z$, per square degree for three survey strategies with line flux limits of $5\times10^{-17}$erg s$^{-1}$cm$^{-2}$, $1\times10^{-16}$erg s$^{-1}$cm$^{-2}$ and $2\times10^{-16}$erg s$^{-1}$cm$^{-2}$ respectively. In each case we show the mean counts and $1\sigma$ uncertainty from the jackknife resampling method, as well as the cosmic variance for the dust model. We present results for two different dust models respectively. The last two rows give the {\it cumulative} counts for surveys with $1<z<2$ and $0.5<z<2$.
Note that the observational efficiency is not included here.}
\begin{tabular}{cccccccc}
\hline
\multicolumn{2}{c}{ }  &  \multicolumn{3}{c}{Dust model fit at all redshifts} & \multicolumn{3}{c}{Dust model fit at high redshifts} \\
\hline
\multicolumn{2}{c}{Redshift} & \multicolumn{3}{c}{Flux limit [erg s$^{-1}$cm$^{-2}$]} & \multicolumn{3}{c}{Flux limit [erg s$^{-1}$cm$^{-2}$]}\\
From & To & $\left(5\times 10^{-17}\right)$ & $\left(1\times 10^{-16}\right)$ &$\left(2\times 10^{-16}\right)$ & $\left(5\times 10^{-17}\right)$ & $\left(1\times 10^{-16}\right)$ &$\left(2\times 10^{-16}\right)$ \\
\hline
0.5 &  0.6  & $29787\pm2363$  &  $16642\pm1311$  &  $7732\pm610$  & $34725\pm2775$ & $19912\pm1550$  & $9820\pm775$\\
0.6 &  0.7  & $36100\pm2481$  &  $18820\pm1308$  &  $7820\pm591$  & $42560\pm2922$ & $23110\pm1600$  & $10420\pm770$\\
0.7 &  0.8  & $29610\pm1734$  &  $14985\pm879$  &  $5815\pm354$  & $34850\pm2025$  & $18567\pm1088$  & $7802\pm470$\\
0.8 &  0.9  & $29992\pm1859$  &  $14072\pm874$  &  $4920\pm341$  & $35750\pm2205$  & $17942\pm1132$  & $6897\pm459$\\
0.9 &  1.0  & $27007\pm1531$  &  $11812\pm673$   &  $3960\pm261$  & $32662\pm1836$ & $15570\pm892$  & $5527\pm340$\\
1.0 &  1.1  & $29597\pm1828$  &  $12120\pm759$  &  $3667\pm242$   & $36517\pm2225$ & $16235\pm1013$  & $5340\pm337$\\
1.1 &  1.2  & $27627\pm1595$  &  $10620\pm622$  &  $2865\pm182$  & $34552\pm1983$ & $14312\pm847$   & $4260\pm266$\\
1.2 &  1.3  & $20987\pm1646$  &  $7537\pm607$  &  $1990\pm172$  & $26615\pm2061$  & $10377\pm840$  & $3055\pm262$\\
1.3 &  1.4  & $24215\pm1550$  &  $8217\pm521$  &  $1835\pm128$  & $31170\pm2009$  & $11615\pm740$  & $3025\pm200$\\
1.4 &  1.5  & $20185\pm1309$  &  $6410\pm454$  &  $1327\pm115$  & $26380\pm1708$  & $9270\pm632$   & $2247\pm172$\\
1.5 &  1.6  & $16742\pm1116$  &  $4790\pm345$  &  $800\pm59$  & $22125\pm1442$    & $7190\pm505$   & $1355\pm100$\\
1.6 &  1.7  & $14670\pm869$  &  $3887\pm254$   &  $667\pm56$  & $19727\pm1144$    & $6075\pm373$   & $1050\pm78$\\
1.7 &  1.8  & $12645\pm848$  &  $2997\pm212$   &  $542\pm54$ & $17817\pm1174$    & $4705\pm322$    & $832\pm70$\\
1.8 &  1.9  & $8095\pm661$  &  $1640\pm140$   &  $310\pm34$  & $11732\pm957$   & $2792\pm233$    & $477\pm51$\\
1.9 &  2.0  & $6792\pm388$  &  $1367\pm96$    &  $255\pm22$  & $9955\pm547$   & $2232\pm139$   & $390\pm32$\\
\hline
1.0 &  2.0  & $18155\pm924$ & $5958\pm303$ & $1426\pm75$ & $23659\pm1201$  & $8480\pm432$ & $2203\pm114$\\
0.5 &  2.0  & $22270\pm1125$ & $9061\pm461$ & $2967\pm156$  & $27809\pm1404$ & $11993\pm610$  & $4166\pm217$ \\
\hline
\end{tabular}
\label{tab:Halpha}
\end{table*}

\section{Predictions for [OIII] emission line galaxies}

\subsection{Comparison with observations}

At redshift $z>2$, both WFIRST and Euclid are able to detect the [OIII] emission lines. Therefore it is necessary to forecast the number counts and redshift distributions of [OIII]-selected galaxies given the survey strategies. 

\begin{figure*}
\begin{center}
\includegraphics[width=8.5cm]{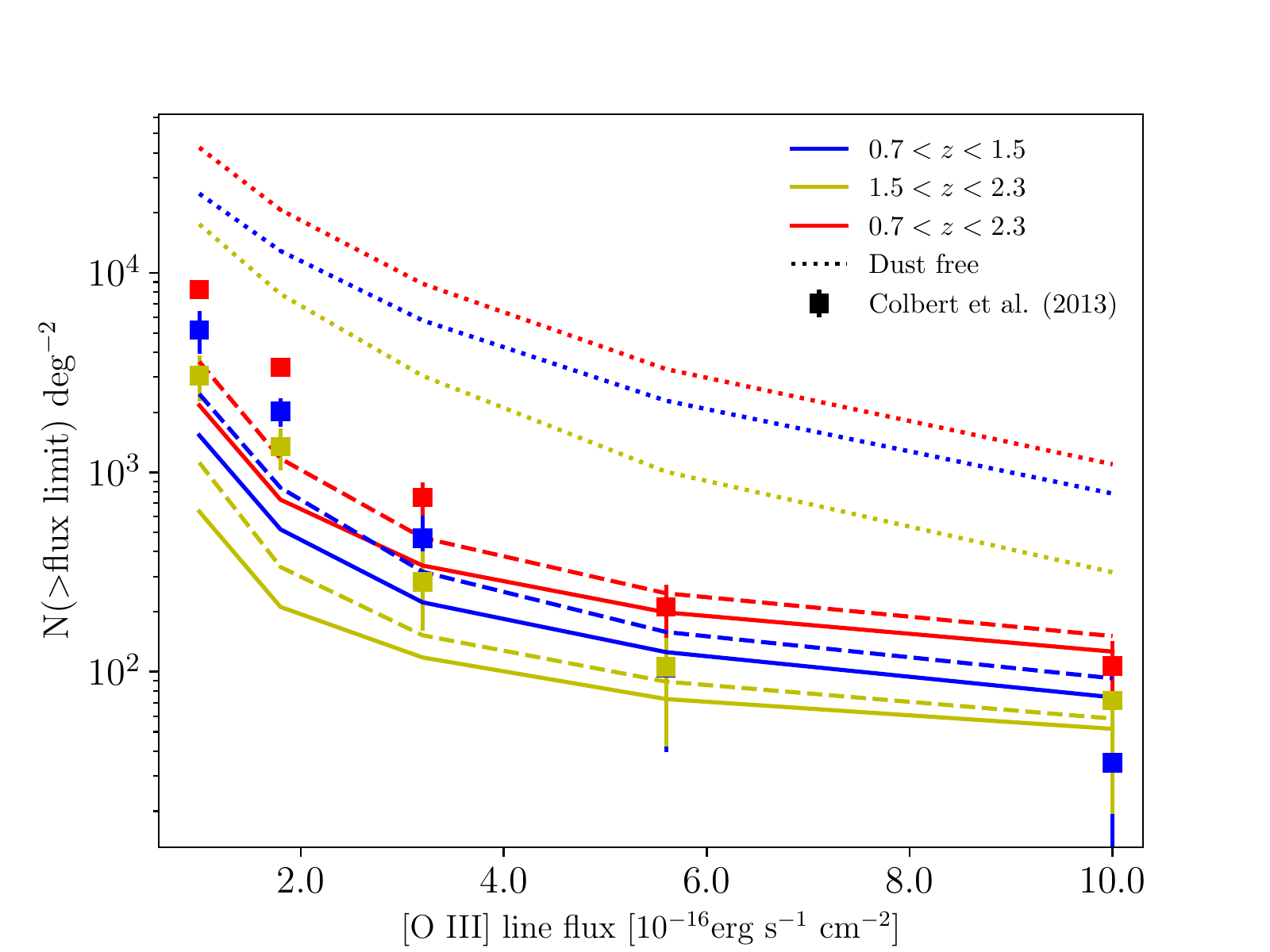}
\includegraphics[width=8.5cm]{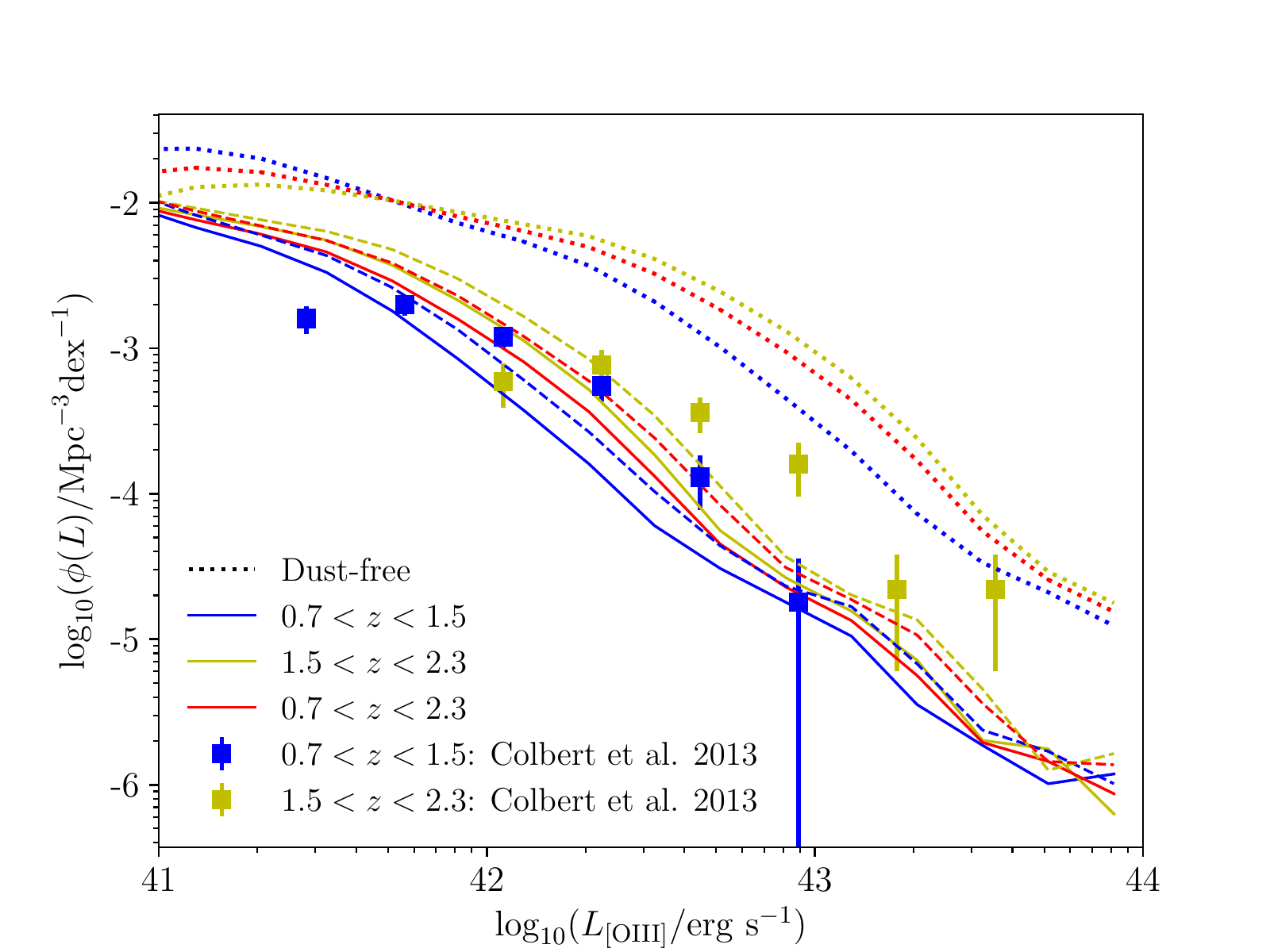}
\caption{Comparison of [OIII] emitting galaxies as predicted by Galacticus and as currently observed.  $Left:$ cumulative number counts with different flux limits in the redshift ranges of \citet{Colbert_2013}. The dotted line is the dust-free result, while the solid line applies our dust attenuation model fitted with the optical depths at all HiZELS redshifts, and the dashed line adopts the average optical depth at high redshift only. The observational data is based on WISP observations. $Right:$ The luminosity function of [OIII] compared with WISP measurements (squares). Both the dust-free (dashed) and dust-attenuated (solid and dashed) results are shown. The uncertainty in our calculation is ignored for plotting purpose.}
\label{fig:Ocounts}
\end{center}
\end{figure*}

Figure \ref{fig:Ocounts} compares our calculation with current observations. Based on the WISP program with 29 fields observed using the G102 and G141 grism, \citet{Colbert_2013} presented the number counts and luminosity function for both H$\alpha$ and [OIII] emitters. We plot their results for the [OIII] emitters on top of our analysis in the same redshift ranges. The left hand panel of Figure \ref{fig:Ocounts} shows the number counts as a function of flux limit. Similar to the result for H$\alpha$ emitters (Figure \ref{fig:Hcounts}), the dust model has a significant effect. Our result is roughly consistent with \citet{Colbert_2013}. However, we notice discrepancies at some flux limits. With the WFIRST flux limit of $\sim1\times10^{-16}$erg s$^{-1}$cm$^{-2}$, our calculation shows a lower number count than \citet{Colbert_2013} especially for the low-redshift galaxies. This can be partially attributed to the dust model we employed in the analysis. At the Euclid flux limit or higher (e.g. $3\times10^{-16}$erg s$^{-1}$cm$^{-2}$), our model predicts similar numbers of galaxies as the WISP measurement, especially for redshift $z>1.5$. This indicates that our analysis can be extrapolated to provide reasonable prediction for WFIRST and Euclid at high redshifts.

The right panel of Figure \ref{fig:Ocounts} presents the luminosity function of [OIII] ELGs as well as the comparison with WISP measurements (\citealt{Colbert_2013}). This result provides further evidence of the consistency between the two. The luminosity function of our [OIII] prediction also has a Schechter-like shape and similar amplitude as the observations. The discrepancy is primarily seen in the lowest luminosity bin for both redshift ranges. This indicates a possible incompleteness in the WISP analysis which can be caused by the number of misidentified single-line [OIII] ELGs at the faint end (\citealt{Colbert_2013}), thus this point was not used in their Schechter fit or further analysis. Excluding this point, we find our luminosity function is comparable with the WISP measurement. 

The results shown in Figure \ref{fig:Ocounts} shows that the number distribution of the [OIII] ELGs is roughly consistent with observations to date. This further validates our galaxy formation and dust models. Together, they form a physically motivated and observationally consistent model useful for future work on galaxy formation and evolution. 

\subsection{Predictions for future surveys}

\begin{figure*}
\begin{center}
\includegraphics[width=8.5cm]{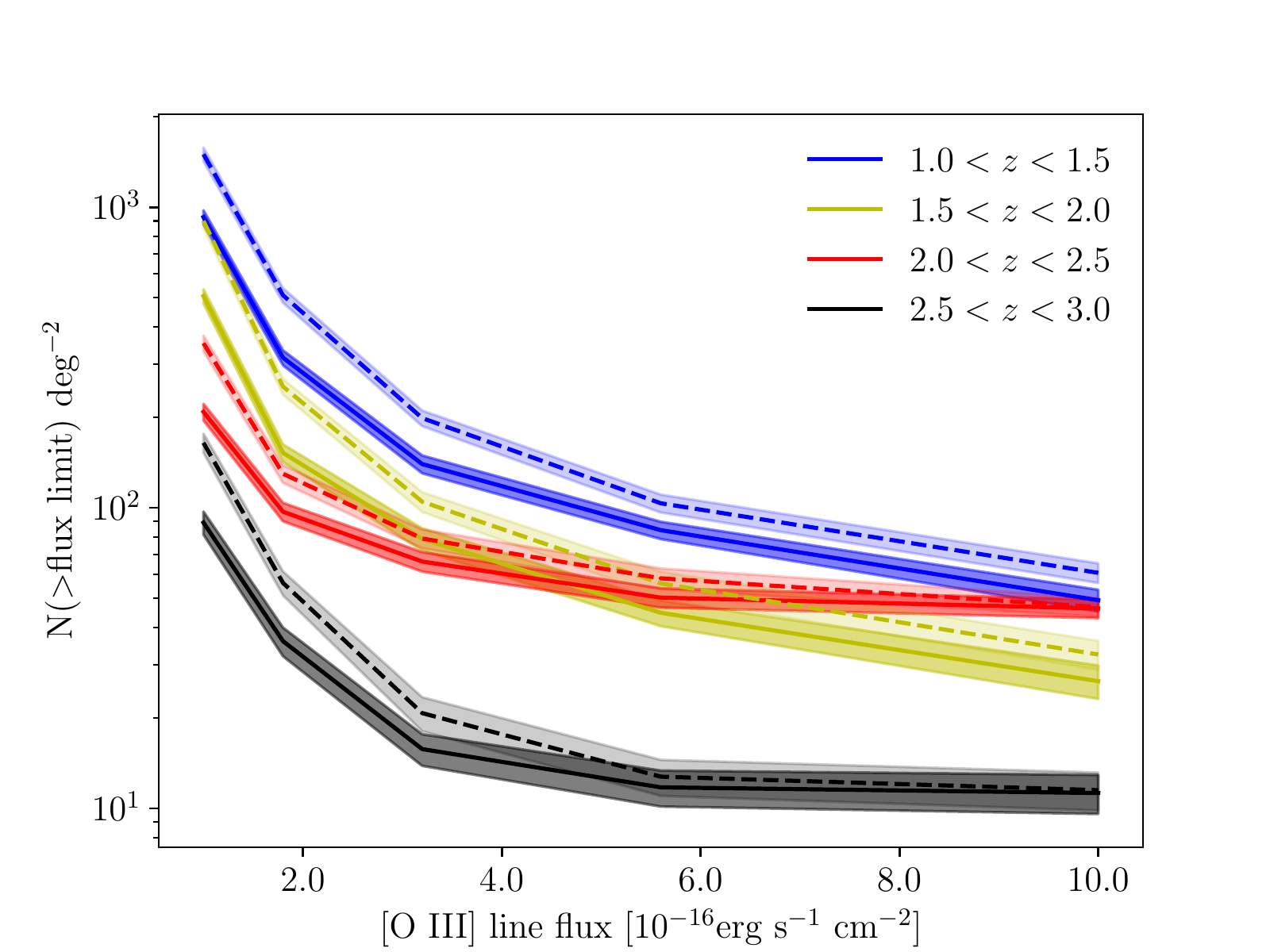}
\includegraphics[width=8.5cm]{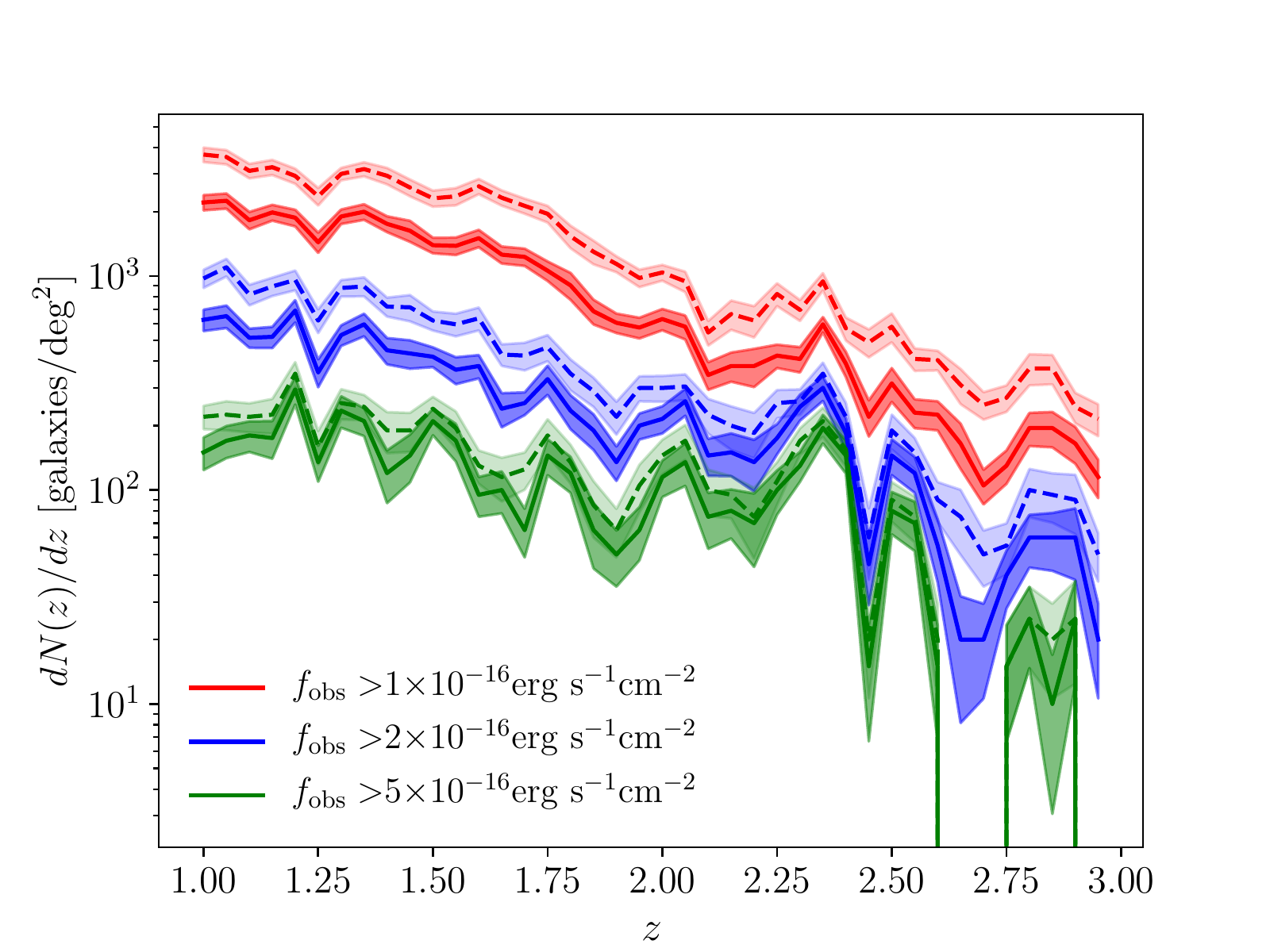}
\caption{Prediction of [OIII] emission line galaxies in the redshift range $1<z<3$. $Left:$ the cumulative flux counts as a function of flux limit for different redshift ranges as shown in the legend. $Right:$ Redshift distribution of the number density of the [OIII] emitting galaxies. The results are shown for different flux limits. The errors are estimated from jackknife resampling method and cosmic variance. The solid line assumes an average optical depth fit to the entire HiZELS redshift range, while the dashed line just uses the high redshift measurements.}
\label{fig:Opre}
\end{center}
\end{figure*}

We present the forecast for [OIII] ELGs in Figure \ref{fig:Opre} and Table \ref{tab:OIII}, in the same format and flux limits as Figure \ref{fig:Hredshift} and Table \ref{tab:Halpha}. The left panel shows the number counts as a function of flux limit within different redshift ranges, and the right hand panel shows the redshift distribution of the number density with different flux limits for different dust models. From this result, we can see that the flux limit is an important factor in shaping the survey strategy. A lower flux limit can enable detection of more faint galaxies, which reflects the fact that the number density distribution of the galaxies or dark matter halos obeys a Schechter-like function. Thus WFIRST will have a denser sampling of the [OIII] galaxies than Euclid, similar to the result for H$\alpha$. With the WFIRST flux limit of $1\times10^{-16}$erg s$^{-1}$cm$^{-2}$, we find that the number density of [OIII] galaxies does not decline significantly with increasing redshift. From redshift 1 to 2.5, the number density with unit redshift is several thousand per square degree. Thus the [OIII] emitters will increasingly dominate the observed galaxy distribution at high redshifts. Combined with the large survey area, it is expected that the observation of [OIII] galaxies will provide accurate measurements of large scale structure at redshifts up to 3. In particular, it is possible that the clustering signal can have comparable significance and accuracy as the H$\alpha$ galaxies at lower redshifts, which can provide important information for the assembly history of galaxies when the universe is only $\sim2$ Gyrs old . 

In addition, the redshift distribution of [OIII] galaxies shown in the right panel of Figure \ref{fig:Ocounts} shows a noticeable decline at high redshift for bright galaxies. This is also revealed by the result for the $2.5<z<3.0$ galaxies in the left hand panel. This can potentially have an impact on the statistical properties of massive galaxies or even galaxy clusters at this redshift. 

\begin{table*}
\centering
\caption{Predicted redshift distribution of the number counts of [OIII]-emitting galaxies, ${\rm d}N/{\rm d}z$, per square degree for three survey strategies with line flux limits of $5\times10^{-17}$erg s$^{-1}$cm$^{-2}$, $1\times10^{-16}$erg s$^{-1}$cm$^{-2}$ and $2\times10^{-16}$erg s$^{-1}$cm$^{-2}$ respectively. In each case we show the mean counts and $1\sigma$ uncertainty from the jackknife resampling method, as well as the cosmic variance for the dust model. We present results for two different dust models respectively. The last two rows show the {\it cumulative} counts for surveys with $1<z<2$ and $2<z<3$ respectively. Note that the observational efficiency is not included here.}
\begin{tabular}{cccccccc}
\hline
\multicolumn{2}{c}{ }  &  \multicolumn{3}{c}{Dust model fit at all redshifts} & \multicolumn{3}{c}{Dust model fit at high redshifts} \\
\hline
\multicolumn{2}{c}{Redshift} & \multicolumn{3}{c}{Flux limit [erg s$^{-1}$cm$^{-2}$]} & \multicolumn{3}{c}{Flux limit [erg s$^{-1}$cm$^{-2}$]}\\
From & To & $\left(5\times 10^{-17}\right)$ & $\left(1\times 10^{-16}\right)$ &$\left(2\times 10^{-16}\right)$ & $\left(5\times 10^{-17}\right)$ & $\left(1\times 10^{-16}\right)$ &$\left(2\times 10^{-16}\right)$ \\
\hline
1.0 &  1.1  &  $6880\pm426$  & $2117\pm140$  & $610\pm58$  & $10437\pm617$ & $3440\pm230$ & $977\pm74$\\
1.1 &  1.2  &  $6590\pm400$  & $1942\pm131$  & $557\pm47$ & $10212\pm589$ & $3167\pm204$ & $922\pm72$\\
1.2 &  1.3  &  $5362\pm387$  & $1562\pm132$  & $427\pm48$ & $8360\pm609$ & $2560\pm191$ & $697\pm67$\\
1.3 &  1.4  &  $6442\pm388$  & $1967\pm128$  & $572\pm45$ & $10027\pm616$ & $3095\pm189$ & $875\pm64$\\
1.4 &  1.5  &  $5752\pm394$  & $1677\pm139$  & $482\pm49$ & $8987\pm578$ & $2732\pm196$ & $752\pm67$\\
1.5 &  1.6  &  $5155\pm328$  & $1350\pm99$  & $340\pm33$ & $7900\pm484$ & $2355\pm163$ & $562\pm50$\\
1.6 &  1.7  &  $5377\pm335$  & $1315\pm93$  & $290\pm29$ & $8365\pm495$ & $2355\pm150$ & $490\pm45$\\
1.7 &  1.8  &  $4362\pm286$  & $1110\pm77$  & $310\pm37$ & $6957\pm460$ & $1930\pm134$ & $455\pm44$\\
1.8 &  1.9  &  $3252\pm266$   & $682\pm66$  & $167\pm22$ & $5412\pm415$ & $1305\pm122$ & $262\pm31$\\
1.9 &  2.0  &  $2680\pm161$   & $605\pm40$  & $180\pm17$ & $4442\pm251$ & $1052\pm68$ & $275\pm28$\\
2.0 &  2.1  &  $1962\pm152$   &  $555\pm62$  &  $235\pm32$  & $3055\pm237$ & $875\pm85$ & $302\pm37$\\
2.1 &  2.2  &  $1422\pm160$    &  $345\pm45$  &  $137\pm23$ & $2265\pm243$ & $597\pm69$ & $190\pm30$\\
2.2 &  2.3  &  $1872\pm175$    &  $395\pm46$  &  $175\pm22$ & $3210\pm287$ & $715\pm79$ & $227\pm31$\\
2.3 &  2.4  &  $1845\pm137$    &  $522\pm41$  &  $265\pm26$ & $3170\pm230$ & $820\pm69$ & $302\pm32$\\
2.4 &  2.5  &  $1547\pm114$    &  $267\pm33$  &  $97\pm17$ & $2722\pm219$ & $520\pm54$ & $122\pm21$\\
2.5 &  2.6  &  $1297\pm103$    &  $275\pm32$  &  $120\pm19$ & $2360\pm169$ & $477\pm50$ & $162\pm21$\\
2.6 &  2.7  &  $1067\pm93$    &  $160\pm23$ &   $30\pm8$  & $2090\pm169$ & $312\pm33$ & $67\pm12$\\
2.7 &  2.8  &  $960\pm78$    &  $140\pm19$  &  $35\pm8$ & $1932\pm149$ & $295\pm33$ & $60\pm11$\\
2.8 &  2.9  &  $1045\pm106$    &  $195\pm29$  &  $67\pm11$ & $1942\pm187$ & $337\pm46$ & $105\pm20$\\
2.9 &  3.0  &  $625\pm66$    &  $122\pm18$  &  $35\pm11$ & $1337\pm117$ & $222\pm29$ & $67\pm15$\\
\hline
1.0 &  2.0  & $5185\pm258$ &  $1433\pm72$  &  $393\pm20$ & $8110\pm406$ & $2399\pm121$ & $627\pm32$\\
2.0 &  3.0  & $1364\pm74$  &  $297\pm18$  &  $119\pm7$ &  $2408\pm127$ & $517\pm30$ & $160\pm10$\\
\hline
\end{tabular}
\label{tab:OIII}
\end{table*}

\subsection{H$\alpha$/[OIII] Ratio}

\begin{figure*}
\begin{center}
\includegraphics[width=18.0cm]{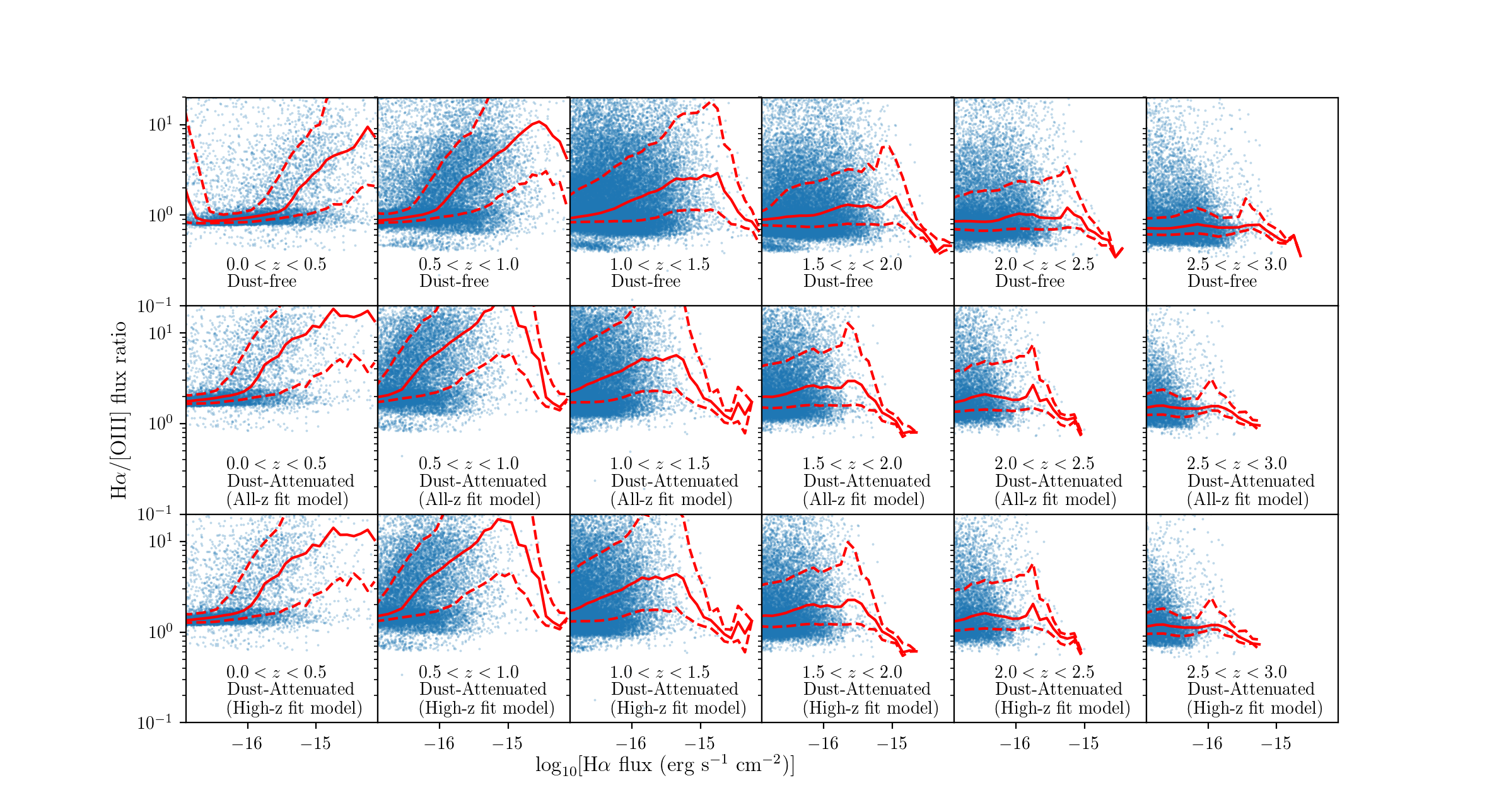}
\caption{Distribution of the flux ratio of H$\alpha$/[OIII] as a function of H${\alpha}$ flux, for both dust-free (top row) and dust-attenuated (middle and bottom row) results. For each panel, we split the galaxy population into small H$\alpha$ bins, and estimate the median (solid line), and $25\%$ and $75\%$ (dashed line) percentile of the flux ratio to represent the overall distribution.}
\label{fig:fluxratio}
\end{center}
\end{figure*}

WFIRST and Euclid can observe both the H$\alpha$ and [OIII] ELGs over a wide redshift range. In order to avoid double counting in the statistical analysis of the galaxy properties, we split the galaxy population at $z=2$. However we note that many of the galaxies can have both emission lines detected and it is worthwhile to examine the relationship between the strength of the emission lines. Therefore we compute the flux ratio of H$\alpha$/[OIII] for the model galaxies from our Galacticus simulation. We split the galaxy population into several redshift bins and present results for both dust-free and dust-attenuated in Figure \ref{fig:fluxratio}.

The top row in the figure shows the dust-free result, which reveals the intrinsic relation between the two emission line fluxes. The middle and bottom rows adopt the dust models described in Section \ref{sec:calibration}. In order to capture the overall trend of the statistics, we split the data in the panel into small bins of H$\alpha$ flux, and plot the median (solid lines)
with the $25\%$ and $75\%$ (dashed lines) percentile on top of the scatter plot. We find that the results show similar patterns of redshift and H$\alpha$ flux dependence of the flux ratio except that the dust extinction brings the overall galaxy population to the faint end. Thus we can expect that the dust model will not significantly impact the observed relation in future surveys. At redshift $z<1.5$, the figure clearly shows the trend of increasing H$\alpha$/[OIII] ratio with H$\alpha$ flux. This is consistent with the analysis based on WISP (\citealt{Colbert_2013}), and earlier analysis reported by \citet{Dominguez_2013}. With increasing redshift, this trend becomes weaker and the ratio approaches a constant across the H$\alpha$ flux distribution. In addition, we find that with increasing redshfit, the [OIII] flux starts to be stronger than H$\alpha$ and dominates the observation. Due to the correlation of the metallicity, stellar mass and luminosity (\citealt{Tremonti_2004, Kobulnicky_2004}), the dependence of the H$\alpha$/[OIII] flux ratio on the galaxy properties can be non-trivial. A thorough investigation of the galaxy formation model using all available observational data is necessary in future analyses.

The result presented in Figure \ref{fig:fluxratio} is a model prediction based on ideal conditions. The only observational effect considered here is the dust extinction model calibrated on the H$\alpha$ luminosity function. Thus we should note the caveat that future observations can be complicated by multiple factors which can affect the H$\alpha$/[OIII] flux ratio distribution. This includes the completeness of the survey, the number of galaxies that have both emission lines detected, the dust extinction model adopted, contamination from other emission lines (e.g. [NII] for H$\alpha$ \citealt{Masters_2014, Masters_2016, Faisst_2016, Faisst_2018}) and so on. 

\section{Discussion and Conclusion}

WFIRST and Euclid missions, as cosmological surveys of the next generation, will play important roles in the investigation of dark energy in the universe by observing a huge cosmic volume over a wide redshift range. Optimizing and evaluating their survey strategy and performance is necessary to forecast the science return. This requires realistic simulation to model the galaxies that they will be able to observe. In this work, we use a semi-analytical galaxy formation model, Galacticus, with a large N-body simulation to predict the number counts and redshift distributions of the H$\alpha$ and [OIII] ELGs for a Euclid-like survey, a WFIRST-like survey, as well as a deeper survey for comparison.

We first calibrate the galaxy formation model and the dust attenuation model to match the observed H$\alpha$ luminosity function. The emission line luminosity of the mock galaxies is computed by the  CLOUDY code. We calibrate the model parameters of Galacticus by exploring the parameter space through a latin-hypercube method. We note that this method is only a simplified and approximate method but can serve as an sufficient modeling for the forecast of galaxy surveys at relevant redshifts. More robust calibration can use method like emulator based on Gaussian Process or other machine learning method (\citealt{Bower_2010}). This can not only find reasonable parameter set for the galaxy formation model based on certain observational datasets, but also enable the examination of possible tensions between the astrophysical processes implemented and thus improve the model building.

In the calibration of the dust model, we employ a method equivalent to the \citet{Calzetti_2000} dust model and measure the optical depth at the H$\alpha$ wavelength to match the observed luminosity function from HiZELS. Then we adopt a dust attenuation model which has a constant optical depth over the entire redshift range or only at high redshift and compare the resulting distribution of our galaxy sample.
This results a physically motivated and observationally consistent model for galaxy population. We then apply this model to process the dark matter halo merger trees from the UNIT simulation and construct a 4 square degree lightcone galaxy catalog. We compare the model prediction of the H$\alpha$ emission line luminosity and number counts with observations from HiZELS and WISP, and find consistent results. Next, we perform an analysis for the [OIII]-emitting galaxies which are also target galaxies of WFIRST and Euclid at redshift $z>2$. The number counts and luminosity function of [OIII] emitting galaxies are found to be roughly consistent with current observations from WISP. We then use Galacticus to extrapolate our prediction to higher redshifts that WFIRST and Euclid can probe and estimate the number density of the galaxies. The results show that at redshift up to 3, the observation of [OIII] galaxies from WFIRST can have a surface density of thousands per square degree, which will significantly extend the redshift reach of WFIRST in probing dark energy.

We examine the relationship between the H$\alpha$ flux and [OIII] flux of the galaxies by computing the flux ratio as a function of H$\alpha$ flux. At low redshift, we find the trend of increasing ratio with H$\alpha$ flux, as reported in earlier studies. The result also approves that galaxies at high redshift are more likely to have [OIII] emission rather than H$\alpha$ emission. In addition, we also notice that applying the dust model doesn't change the overall behavior significantly. This can avoid introducing extra bias in the analysis with observational data. However, we should note that the dust extinction in the real universe can be more complicated than modeled here. 

Although the results presented in this paper are consistent with current observations and earlier work in \citet{Merson_2018}, it should be noted that they are expectations under somewhat ideal conditions where the only observational effect is the dust attenuation. In real observation, the number counts and redshift distribution of the emission line galaxies can be affected by multiple factors, such as the redshift failures, survey completeness, contamination from other emission lines and so on.

The number counts prediction for galaxies is one of the first steps in the assessment of future surveys like WFIRST or Euclid. The analysis of galaxy clustering (correlation function or power spectrum) can be the next step. For instance, \citet{Merson_2019} employ a HOD approach to simulate a WFIRST-like and Euclid-like galaxy survey, measure the clustering and estimate the linear bias. The semi-analytical galaxy formation model, Galacticus we used in this analysis is also able to produce galaxy catalogs with large survey area and accuracy. We will present the clustering analysis based on this approach, including both galaxy correlation function, power spectrum and higher order statistics in a future work.

\section*{Acknowledgements}

We thank the anonymous referee for the valuable comments and suggestions that have helped us improve the contents of this paper. We thank James Colbert for providing the WISP measurements in the analysis, and Alex Merson for helpful discussions. This work is supported in part by NASA grant 15-WFIRST15-0008, Cosmology with the High Latitude Survey WFIRST Science Investigation Team (SIT). 
GY would like to thank MINECO/FEDER (Spains)  for financial support under project grants AYA2015-63810-P and  PGC2018-094975-B-C21.
The UNIT simulations have been done in the MareNostrum Supercomputer at the Barcelona Supercomputing Center (Spain) thanks to the  cpu time awarded by PRACE under project grant number 2016163937. This work used the Extreme Science and Engineering Discovery Environment (XSEDE), which is supported by National Science Foundation grant number ACI-1548562 (\citealt{XSEDE_2014})

\rm{Software:} Python,
Matplotlib \citep{matplotlib},
NumPy \citep{numpy},
SciPy \citep{scipy}


\bibliographystyle{mnras}
\bibliography{emu_gc_bib,software}

\appendix

\section{Galaxy properties from Calibration}

\begin{figure*}
\begin{center}
\includegraphics[width=8.5cm]{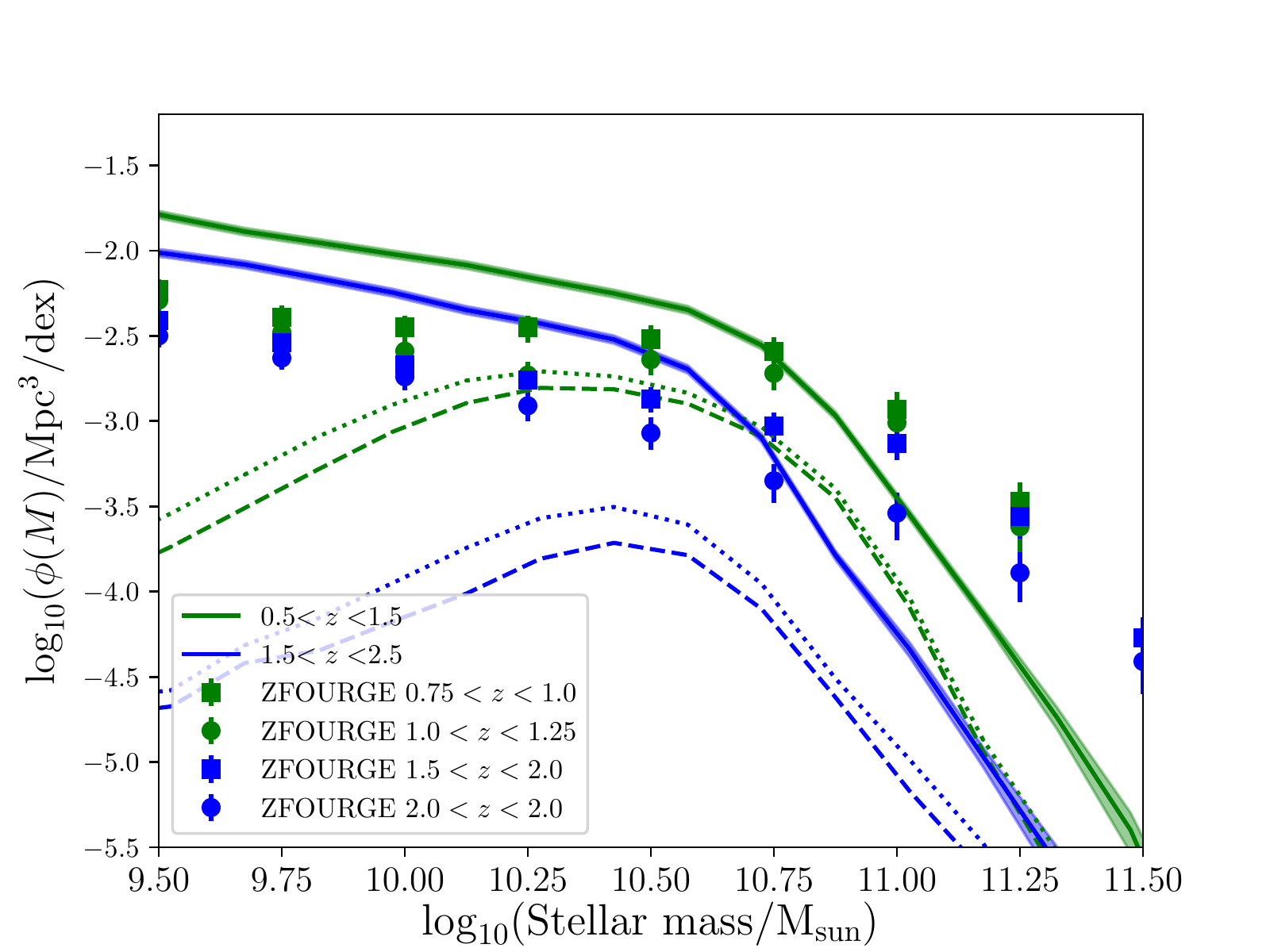}
\includegraphics[width=8.5cm]{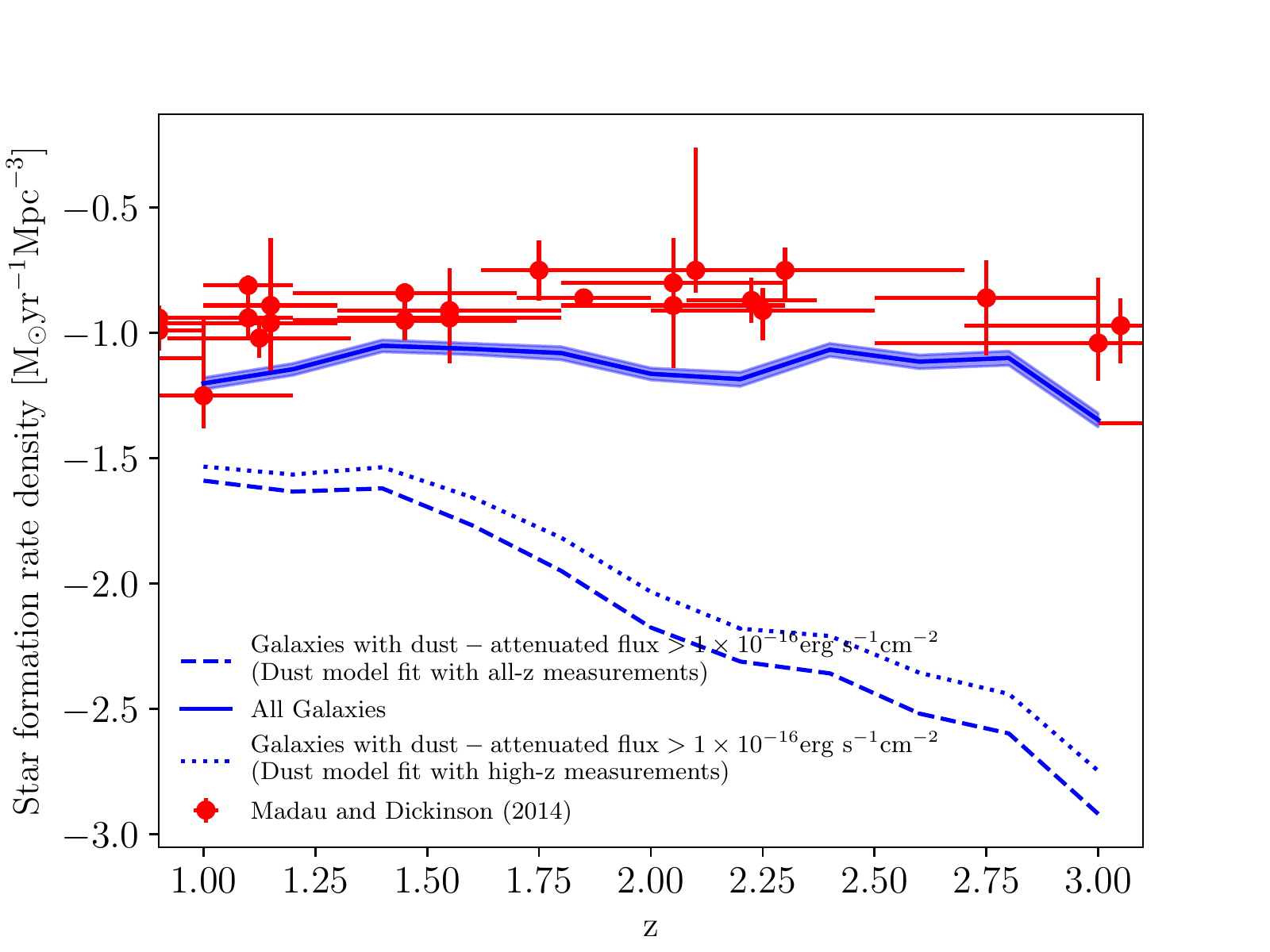}
\caption{$Left:$ The stellar mass function of our mock galaxy catalog. Solid lines are from the raw galaxy, while the dashed and dotted lines are obtained by only selecting galaxies with dust-attenuated H$\alpha$ flux higher than $1\times10^{-16}\text{erg}/\text{s}/\text{cm}^{2}$. For comparison, we also plot the observational measurements from  ZFOURGE (\citealt{Tomczak_2014}), in the similar redshift range. $Right:$ the star formation rate density as a function of redshift $z$. Solid line is for the whole galaxy sample, the dashed and dotted lines correspond to a brighter subsample. The red dots with errorbars are from a compilation of \citet{Madau_2014} in the same redshift range. The shaded area in both panels is the uncertainties from jackknife resampling method and cosmic variance. The dust models used here correspond to the optical depth averaged in the entire HiZELS redshift range (dashed) and only the high-z measurements (dotted).}
\label{fig:SMF}
\end{center}
\end{figure*}

In the left panel of Figure \ref{fig:SMF}, we show the stellar mass function at $z=1$ and $2$ for the whole galaxy catalog and a brighter subsample. The result shows a Schechter-like shape as expected and correct redshift dependence, e.g. the stellar mass function at lower redshift has higher amplitude due to the hierarchical growth of the dark matter structure. After applying the flux cut at $1\times10^{-16}\text{erg}/\text{s}/\text{cm}^{2}$ of H$\alpha$ emission line, most of the less massive galaxies are removed. The remaining galaxies are due to the scatter between the stellar mass and luminosity. For comparison, we also show the measurements from the FourStar Galaxy Evolution Survey (ZFOUGRE, \citealt{Tomczak_2014}) at similar redshift range. Our prediction has roughly consistent amplitude as the observation, but there is discrepancy at low mass and high mass end. We should note that the galaxy model we used is calibrated on the luminosity function measurements, therefore the measurement of stellar mass function has minimal weight in the analysis. The right panel of Figure \ref{fig:SMF} shows the star formation rate density in the redshift range $1<z<3$, as well as a compilation of the latest observational data from \cite{Madau_2014}. The peak value in this redshift range is close but lower than observations. Given the large uncertainties from observations, our result is still consistent within $\sim1\sigma$. In addition, the star formation history model which \cite{Madau_2014} fit to these data predicts a stellar mass density higher than many observations (by around $\sim0.2$ dex on average), which may indicate that star formation rate densities are overestimated observationally. Taking this factor into account would bring our prediction into very good agreement with current observations.

\begin{figure*}
\begin{center}
\includegraphics[width=8.8cm]{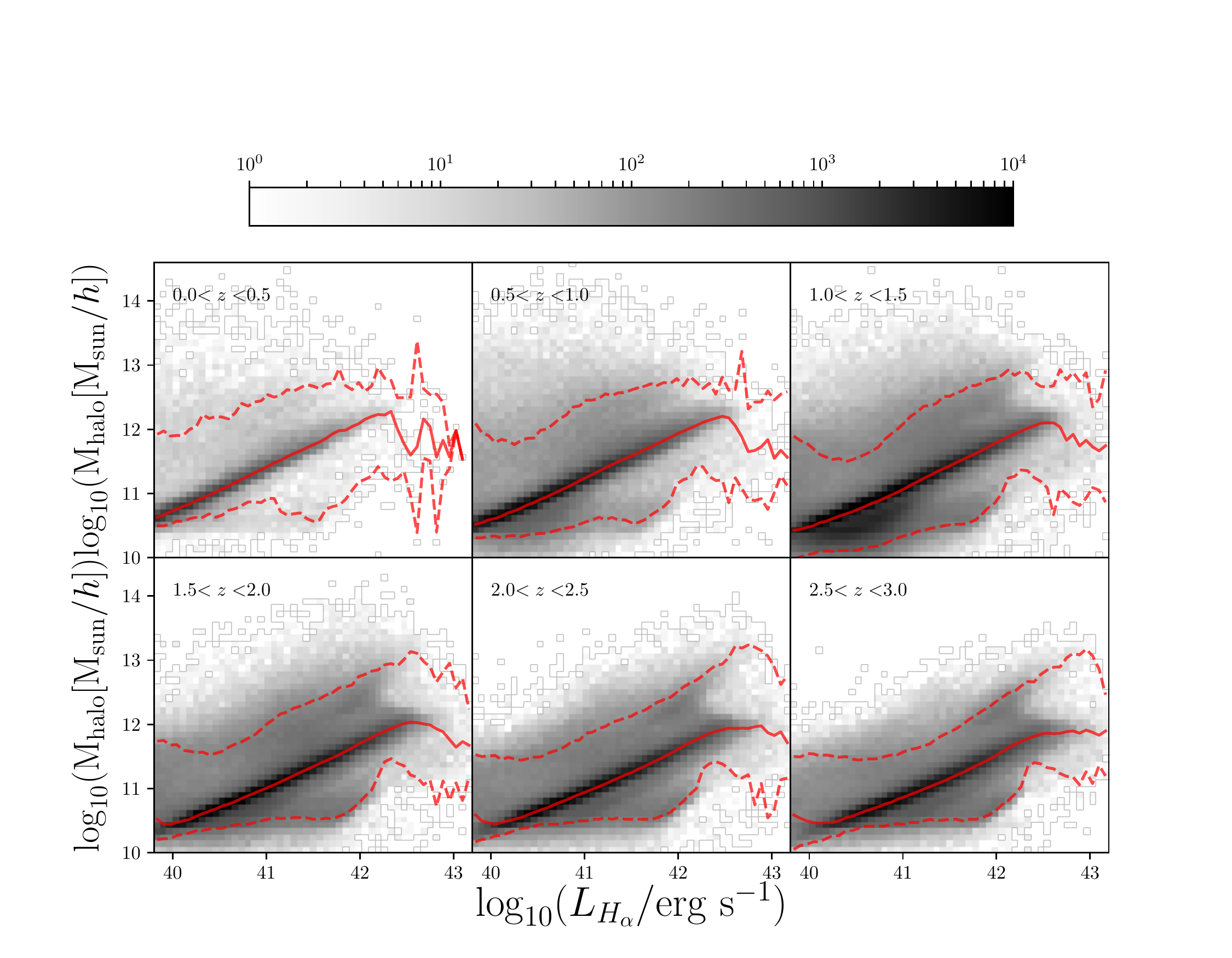}
\includegraphics[width=8.8cm]{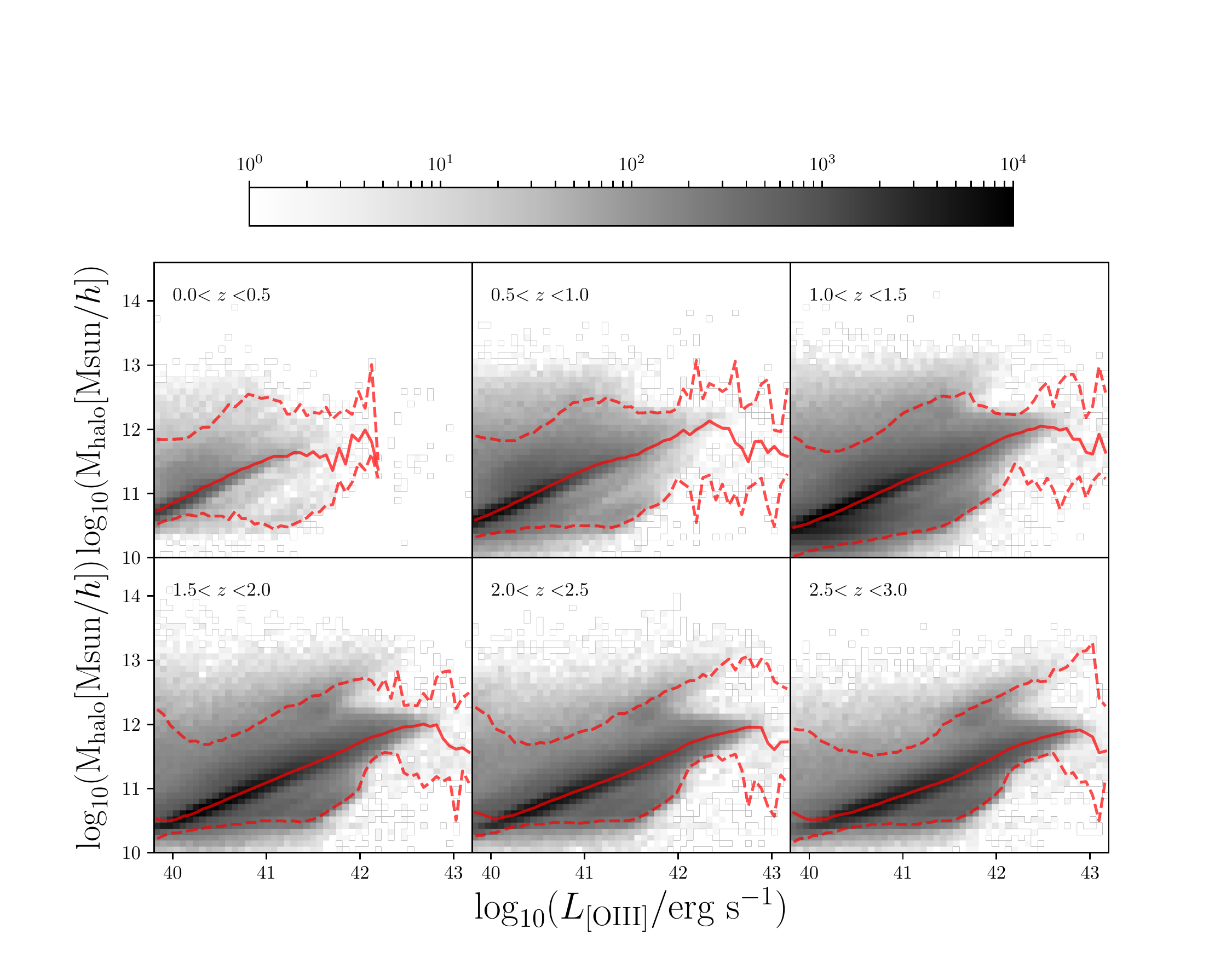}
\caption{The dark matter halo mass as a function of dust-attenuated line luminosities in different redshift ranges (dust model adopts a constant optical depth at the H$\alpha$ wavelength for the high-z measurements). The solid red line represents the median in the distribution, while the dashed red lines are inner 95\% percentile. The gray squares show density distribution of galaxies on the halo mass vs emission line luminosity panel, with deeper gray indicating greater concentration (see the scale on the top). $Left:$ H$\alpha$; $Right:$ [OIII]. }
\label{fig:Hmass}
\end{center}
\end{figure*}

As in \cite{Orsi_2014}, we present the relation between dark matter halo mass and dust-attenuated emission line luminosity in Figure \ref{fig:Hmass}, for H$\alpha$ (left) and [OIII] (right) respectively. We show only results obtained by assuming a constant dust optical depth at the H$\alpha$ wavelength fitted with high-z LF measurements. Results using our alternative dust model are similar. It is clear that there is a strong positive correlation between the emission line luminosity and dark matter halo mass and the relation is close to linear in log space when the emission line luminosity is lower than $10^{42}/\text{erg s}^{-1}$ at redshifts higher than 0.5. Therefore observations of the bright emission line galaxies can provide important information for the massive dark matter halos and their clustering properties.

\bsp	
\label{lastpage}
\end{document}